\documentclass[12pt]{article}
\usepackage[utf8]{inputenc}
\usepackage{xurl}
\usepackage{scicite,times}
\newenvironment{sciabstract}{%
\begin{quote} \bf}
{\end{quote}}

\usepackage{soul}
\usepackage{color,xcolor}
\usepackage{amsmath,mathtools,amsfonts,amssymb,mathptmx}
\usepackage{graphicx}
\usepackage{txfonts}
\usepackage{subcaption}
\usepackage{multirow,xspace}
\usepackage{clrscode3e,algorithm}

\usepackage[authormarkup=none,xcolor=dvipsnames]{changes}

\setaddedmarkup{\textcolor{blue}{#1}}
\setdeletedmarkup{\textcolor{red}{\sout{#1}}}

\usepackage{setspace}
\hypersetup{
    colorlinks,
    linkcolor={red!90!red},
    citecolor={blue!50!black},
    urlcolor={cyan!100!black}
}

\usepackage{url}

\newcommand{\myrange}[3][]{ \ensuremath{#2}-\ensuremath{#3} \ensuremath{#1}}
\graphicspath{{Final_Figures/}}

\title{DeepAutoPIN: An automorphism orbits based deep neural network for 
characterizing the organizational diversity of protein interactomes across the 
tree of life}

\author
{Vikram Singh$^{\dagger}$, Vikram Singh$^{\ast}$\\
\\
\normalsize{Centre for Computational Biology and Bioinformatics, Central
University of Himahcal Pradesh,}\\
\normalsize{Dharamshala,  Himahcal Pradesh, 176206, India} \\
\normalsize{$^\ast$E-mail: vikramsingh@cuhimachal.ac.in}
}

\date{}

\begin{document} 
\baselineskip24pt
\maketitle

%\textbf{Corresponding author:} Dr. Vikram Singh; Assistant Professor; Centre 
%for Computational Biology and Bioinformatics, School of Life Sciences, Central 
%University of Himahcal Pradesh, 176206, India; Contact: +91 9816444313; 
%E-mail: vikramsingh@cuhimachal.ac.in

%\textbf{Classification:} BIOLOGICAL SCIENCES: Biophysics and Computational 
%Biology; Evolution; Systems Biology.
    
\begin{sciabstract}
\section*{Abstract}
The enormous diversity of life forms thriving in drastically different 
environmental milieus involves a complex interplay among constituent proteins 
interacting with each other. However, the organizational principles 
characterizing the evolution of protein interaction networks (PINs) across the 
tree of life are largely unknown. Here we study 4,738 PINs belonging to 16 
phyla to discover phyla-specific architectural features and examine if there are 
some evolutionary constraints imposed on the networks' topologies. We utilized 
positional information of a network’s nodes by normalizing the frequencies of 
automorphism orbits appearing in graphlets of sizes 2-5. We report that orbit 
usage profiles (OUPs) of networks belonging to the three domains of life are 
contrastingly different not only at the domain level but also at the scale of 
phyla. Integrating the information related to protein families, domains, 
subcellular location, gene ontology, and pathways, our results indicate that 
wiring patterns of PINs in different phyla are not randomly generated rather 
they are shaped by evolutionary constraints imposed on them. There exist subtle 
but substantial variations in the wiring patterns of PINs that enable OUPs to 
differentiate among different superfamilies. A deep neural network was 
trained on differentially expressed orbits resulting in a prediction accuracy of $85\%$.

\end{sciabstract}

\textbf{Keywords:} Automorphism orbits, Graphlets, Protein interaction 
networks, Deep neural network, Tree of life.

\textbf{Significance statement}

The organizational principles characterizing the evolution of protein 
interaction networks (PINs) across the tree of life are largely unknown. In this 
study, we have characterized evolutionarily conserved topologies across the 
three domains of life, along with phylum-specific restrictively conserved wiring 
patterns. Using functional information and statistical testing, we deduced that 
the evolutionary constraints had shaped the wiring patterns of PINs across the 
tree of life.

\section*{Introduction}

Life manifests itself in fundamentally diverse forms ranging from microbes to 
plants thriving in vastly different environmental milieus 
\cite{castelle2018major}. Underlying this utmost diversity, there exists a 
universal cellular machinery that is governed by a population of small, 
self-organising molecules called proteins. The complex interplay among 
constituent proteins of a cell that interact with each other to control vital 
biological processes, like, metabolic reactions, molecular transport, immunity, 
gene expression, signaling \textit{etc.} produces different phenotypes that 
allow sustenance of life in varying environmental conditions 
\cite{boeckmann2005protein}. The mechanistic details of how evolutionary 
constraints shape the topology of interacting proteins in an organism are still 
elusive despite a large number of sequence-based studies on genomic evolution 
\cite{sorrells2015intersecting}. Recent advancements in high-throughput
technologies have provided a large volume of biological data, including 
protein-protein interactions (PPI), from different domains that can be analyzed 
using the principles of networks science to gain insights into the mechanisms 
underlying these complex systems 
\cite{newman2003structure,snider2015fundamentals,zhang2019multimodal}. 
Fundamental to each network type is the connectivity patterns giving rise to a 
specific topology. The topology of complex networks represents their underlying 
generative phenomena, and it varies across the systems of different disciplines 
(inter-disciplinary) \cite{milo2004superfamilies} as well as within the classes 
of systems belonging to a particular discipline (intra-disciplinary) 
\cite{newaz2020network}. For example, the topological characteristics of 
different types of biological networks, like, Protein interaction networks 
(PINs), residue interaction networks (RINs), food webs, brain networks 
\textit{etc.} are different.

Inter-disciplinary networks, such as PINs, can further be sub-categorized at any 
level of taxonomic rank such as phyla, order, class \textit{etc.} 
(intra-disciplinary networks). This variability is due to the existence of 
inherent similarities along with unique characteristic features among networks 
belonging to different classes \cite{ikehara2017characterizing}. PINs are 
conspicuous as they provide structural and functional elucidation of 
interacting proteins. Much research has been devoted to application of either 
single node properties, like, various types of node centrality measures 
\cite{ikehara2017characterizing,albert2002statistical} called microscopic 
structural features or global network properties, like,  small-world, 
scale-free, average path length \cite{rossi2019complex} called macroscopic 
structural features to characterize the topological diversity of PINs and other 
classes of networks. However, delving deeper into the networks' structures 
shows that there are small assemblages of subsystems which are functioning 
independently. Although these subsystems have their own functions, their 
amalgamation give rise to complex systems behavior 
\cite{milo2002network,alon2007network}. So, deciphering these interconnection 
patterns is very crucial to gaining insights into the system's structure and 
function \cite{prvzulj2007biological}.

In the past few decades, attempts have been made to compare PINs with other 
classes of networks and cluster different networks into their respective 
classes using mesoscopic structural features, like, modularity, motifs and 
graphlets, resulting from network properties computed on groups of nodes, which 
offer much higher topological information 
\cite{milo2004superfamilies,ikehara2017characterizing,prvzulj2004modeling}. 
Local sub-structures of a large complex network, namely, the network motifs and 
graphlets are core elements of its constituent design, which impose constraints 
and govern many aspects of the emergent systems dynamics \cite{milo2002network}. 
Network motifs, that are over-represented, partial, small subgraphs, have been 
used for inter-disciplinary network characterization and clustering 
\cite{benson2016higher}. Graphlets that are defined as small non-isomorphic 
subgraphs and are induced on large networks, have also been used extensively to 
identify network commonalities and alignment 
\cite{prvzulj2007biological,kuchaiev2010topological,malod2017unified, 
liao2009isorankn}. Similarly, communities, that are the densely connected 
subgroups of nodes with very sparse edges among the members of other subgroups, 
have also been implemented to differentiate inter-disciplinary classification 
of various types of networks \cite{onnela2012taxonomies}.

Although a large volume of research is available on exploiting PIN's 
architecture, however, most of the studies have been performed on either 
intra-disciplinary pairwise network comparison of very few PINs or 
inter-disciplinary comparison with networks belonging to other categories. A 
systems scale extensive study leveraging topological diversity to characterize 
PINs across three domains of life or any other biological taxonomic hierarchy is 
still lacking. Here we study $4,738$ PINs belonging to $16$ phyla from three 
domains of life to characterize the organizational diversity underlying them by 
utilizing normalized frequencies of automorphism orbits appearing 
in graphlets of sizes 2-5. We report that orbit usage profiles (OUPs) of 
networks are contrastingly different across the three domains of life as well 
as at the scale of phyla. Integrating the biological information of five 
functional annotations we found that wiring patterns of PINs in different phyla 
shaped by evolutionary constraints imposed on them.
    
\section*{Materials and Methods}

\subsection*{Network Data}

All the available $5,090$ protein interaction networks (PINs) in STRING (v11) 
\cite{szklarczyk2016string} were denoted as PIN data set. This data set was 
classified into three domains of life (superfamilies) viz. Archaea, Bacteria 
and Eukaryota using NCBI's Taxonomy database \cite{schoch2020ncbi}. In any 
individual network, we considered only interactions having STRING confidence 
score $700$ or above. Further, using the taxonomy database, second data set of 
PINs was created by dividing the networks at phylum rank in the biological 
taxonomy. All the phyla with $30$ or more networks, resulting in $C = 16$ 
categories, were considered for further studies.

\subsection*{Random network models}

Two ensembles of random networks based on Erd\H{o}s-R\'{e}nyi (ER) and density 
dependent scale-free (DDSF) algorithms were generated corresponding to each 
network belonging to $16$ phyla considered here. ER networks ($G(n,m)$) 
with same number of nodes $n$ and links $m$ corresponding to real networks were 
generated by randomly selecting a set of $m$ edges among $\binom{ \binom{n}{2} 
}{m}$ possible edges \cite{erdos1959On}. Since the PINs used in this study 
have varying densities so we modified the original Barab\'{a}si-Albert (BA) 
model of stochastic, scale free growth \cite{barabasi1999emergence} in such a 
way that it accounts for density of the networks. This method maintains both 
the essential ingredients of BA model that are growth and preferential 
attachment, however, the growth is now dependent on density of the real 
network. A random network $H^{\prime}(n^{\prime}, m^{\prime} )$ corresponding 
to real network $(H(n, m))$ is generated by introducing new nodes with degree 
$k$ or $k + 1$ such that $k$ is a function of average degree of the real 
network. The algorithm completes in two steps, namely, (\romannumeral 1) 
generation of the seed network, and (\romannumeral 2) extension of the seed 
network. Initially the seed network contains a single edge between two nodes 
thus have \(n^{\prime} = 2,\ m^{\prime} = 1\) and grows by consecutively adding 
new nodes with degree $k = n^{\prime}-1$ unless the order of seed network 
becomes larger than average degree of real network \(\langle k \rangle = 
\frac{2m}{n}\).  After the construction of seed, subsequent nodes are added to it 
consecutively where each incoming node connects to $k$ or $k + 1$ number of 
nodes such that \(k \leq \langle k_{rem} \rangle \leq k +  1\){}, with 
probabilities \(k + 1 - \langle k_{rem} \rangle\) and \(\langle k_{rem} \rangle 
- k\) respectively. \(\langle k_{rem} \rangle = \frac{2(m - m^{\prime})}{n - 
n^{\prime}}\) is the average degree of the remaining network and is recomputed 
for every node added to growing network (Algo \ref{Algo:A1}).

\subsection*{Orbit Usage Profile (OUP)}

Orbits are sets of symmetrically equivalent nodes present in each graphlet 
\cite{hovcevar2014combinatorial}. In graphlets of order up to $5$ nodes, there 
exist $73$ sets of automorphism groups. To define OUP, first the number of 
occurrences $(o_{ij})$ of the $j^{th}$ node in $i^{th}$ orbit, where $i \in 
\{0,...,72\}$, were enumerated for all the real networks. This $73$ dimensional 
orbit count vector characterizes the local neighborhood of 
each node in the network. Since large sized orbits inherently contain some of 
the smaller orbits as well as the orbit itself, for example the orbit $2$ 
contains orbit $0$ and itself, therefore a redundancy is introduced in the orbit 
count. To deal with this, a weight $w_i$ is assigned to the $i^{th}$ orbit by 
counting the number of other orbits present in the $i^{th}$ orbit and the orbit 
itself denoted as $(w_i)$ that is further normalized as follows \[w_i = 1 - 
\frac{log (o_i)}{log (73)}\] Greater the value of $w_i$, more the importance of 
the orbit as it is less redundant \cite{milenkovic2008uncovering}. Finally, 
weight adjusted relative orbit frequency (WAROF) for a network $G(n,m)$ of 
order $n$ and size $m$ was computed for every orbit $i$ as $F_i(G)=\frac{w_i 
N_i(G)}{T(G)}$, where $N_i(G) = \sum \limits_{j=1}^n o_{ij}$ is the total orbit 
count for the $i^{th}$ orbit in all nodes of network $G$ and, $T(G)=\sum 
\limits_{i=0}^{72} {N_i(G)}$ is total orbit count of $G$ 
\cite{prvzulj2004modeling}. In this way, WAROFs were calculated for all the 
$73$ orbits and the resulting  $73$ dimensional vector represents an orbit 
usage profile (OUP) corresponding to each network. Similarly, OUPs were 
computed for each of the networks in the ER and DDSF ensembles.

\subsection*{Differential orbits expression}

To identify diffferentially expressing orbits, we first took all the OUPs 
belonging to a phylum and then for every orbit $i$, the mean and standard 
deviation of its counts were calculated that were  used to further compute the 
\textit{Z-statistic} for $j^{th}$ network as follows \[Z_{ij} (X)= \frac{o_{ij} 
- \bar o_i}{ {\sigma }_{o_i}}\] $\bar o_i$ and ${\sigma}_{o_i}$ are the mean 
and standard deviation of $i^{th}$ orbit of the PINs of species belonging to 
phylum $X$. Orbits for which \romannumeral 1) $Z score$  $\geq 2.58$ for both 
ER and DDSF, and \romannumeral 2) the condition (\romannumeral 1) was satisfied 
by at least $95\%$ of networks belonging to that phylum were considered as 
diffferentially expressed orbits (DEO). We then averaged orbit counts of 
differentially expressing orbits for every phylum and pairwise distance 
correlation between these phyla was computed. The advantage of using distance 
correlation \cite{szekely2007measuring} as similarity index is mainly due to 
its ability to measure non-monotonic relationship between two variables. Two 
variables $X$ and $Y$ are independent if $\mathcal{R} (X, Y) = 0$, and a value 
of $1$ implies equal linear subspaces of the variables. It is defined as \[ 
\mathcal{R} (X, Y) = \frac{dCov(X,Y)}{\sqrt{dVar(X) \ dVar(Y)} }\] 
Where the distance covariance is defined as $dCov^{2} =  \frac{1}{C^2} \sum 
\limits_{i,j} x_{i,j} . y_{i,j}$. $x_{i,j}$ and $y_{i,j}$ are the $(i,j)^{th}$ 
elements of double centered distance matrices $\mathbf{X}$ and $\mathbf{Y}$ 
computed in an $73$ dimensional Euclidean space.

\subsection*{Functional validation}

To further assess the functional similarity between wiring patterns of different 
species, we constructed $16$ sets, each containing $30$ networks selected 
uniformly at random, from every phyla and identified orbit proteins touching 
$57$ differentially expressed orbits by reverse mapping orbit frequency 
information to network data. Orbit proteins (set of proteins touching the 
respective orbit) for $i^{th}$ orbit of $j^{th}$ network in $k^{th} (k = \{1 
... 16\})$ phylum were identified and then pooled all the orbit proteins of 
$i^{th}$ orbit from all the networks in $k^{th}$ phyla and constructed $57$ 
sets of orbit proteins each representing a differentially expressed orbit (DEO) 
for every phyla. We then obtained five functional annotations namely, gene 
ontology (GO) terms, pathways, protein domains, protein family and subcellular 
localization (SL) corresponding to each orbit protein from UniProt KB 
\cite{apweiler2004uniprot}. It is based on the premise that PPIs having similar 
wiring pattern are evolutionarily conserved across species and hence should 
have high similarity score for PINs within phylum and low similarity values 
otherwise. Furthermore, between every pair of phyla we computed pairwise Jaccard 
Coefficient (JC) among all the DEOs ($80 \%$ orbit protein annotation coverage) 
and obtained their average to construct a $16 \times 16$ average similarity 
matrix for all but SL functional annotations. Since loosely connected regions 
(low degree nodes) in the network represent substantial lack of PPIs 
information in those regions, so functional information encoded from these 
proteins may also be incomplete that may lead to false conclusions. In order to 
avoid this situation and draw consistent patterns across species, we introduce 
a concept of orbit protein annotation coverage (OPAC). For that, first the 
number of orbit proteins covered by a particular functional annotation term 
were enumerated. Top $80\%$ of the orbit proteins that are covered by at least 
one functional annotation term (\textit{i.e.} OPAC 80\%) were selected for 
further analysis.

\subsection*{Preprocessing of OUPs and classification into different phyla}

Each of the above computed $D-dimensional$ feature vector (OUP) from $16$ phyla 
are used for the network classification. The input data was split into training 
($80\%$) and test ($20\%$) sets stratified according to proportion of each 
class in the complete data set. Every feature of the two sets was 
then standardized to have $\mu = 0$ and $\sigma = 1$.  A deep neural network 
$f$ (DeepAutoPIN) consisting of one input, one softmax and four dense hidden 
layers, comprising of $100, 160, 160$ and $60$ units, respectively, was learned 
using 5-fold cross validation approach. The hyper-parameters, rectified linear 
unit (relu) activation function, stochastic gradient-based adam optimizer and 
$l2=0.05$ regularization were used to train the classifier.  Class label of 
each unknown network in held out $20\%$ of the test data was then predicted 
using multiclass classifier $f$. 

\section*{Results and Discussion}

From a total of $5,090$ PINs in STRING v11, we could associate $5,007$ networks 
with $71$ phyla by leveraging taxonomic information from NCBI taxonomy database 
\cite{schoch2020ncbi}. Since, a large number of phyla were having very 
few networks so to ensure that the classification task retain sufficient 
statistical power a threshold of $30$ networks per phyla was set and only 
phylum having $30$ or more networks were considered for further study 
\cite{figueroa2012predicting}. This results in $16$ phyla including $4,738$ 
networks among which $162$ are Archaeal, $4,213$ are Bacterial and $363$ are 
Eukaryotic networks (Table S1). Small subgraphs have been used extensively to 
describe a node's local neighborhood \cite{prvzulj2004modeling,alon2007network}, 
its position in a network and quantify the topological similarity between 
different networks \cite{prvzulj2007biological,milo2004superfamilies}. Their 
applications are based upon the basic premise that wiring patterns of proteins 
involved in a particular function are conserved to a large extent, suggesting 
that proteins having similar wiring patterns (topological signatures) may have 
similar physiological properties 
\cite{davis2015topology,milo2004superfamilies}. Several graph based approaches 
leveraging conserved wiring patterns of proteins to establish relationship 
between network topology and protein function have already been successfully 
used \cite{samanta2003predicting,clark2014comparison}. A basic assumption made 
by all these studies is that the wiring patterns of annotated proteins for each 
function are similar which, however, may vary due to varying effects of 
evolution on different parts of the networks \cite{davis2015topology}. 
Therefore, essential functions that are vital for organisms growth and 
development may experience evolutionary constraints on their wiring pattern. 
However, functions that are more specific to species may experience less 
conservative forces on their topology and their wiring pattern can vary. 
Thus in this study we are proposing a more robust approach to identify most 
prominent topological patterns conserved across different species. Orbit count 
distributions corresponding to every node of a network were enumerated and were 
further normalized (see Methods) to a single $73$ dimensional feature vector 
called orbit usage profile (OUP) that is representative of whole network. Same 
process is repeated for every PIN resulting in $4,738$ profiles for real 
networks and same number of OUPs for corresponding ER and DDSF based random 
networks. 

\subsection*{Orbit enrichment and differential orbit expression analysis}
\begin{figure}[h]
    \centering

    \includegraphics[width=\textwidth]{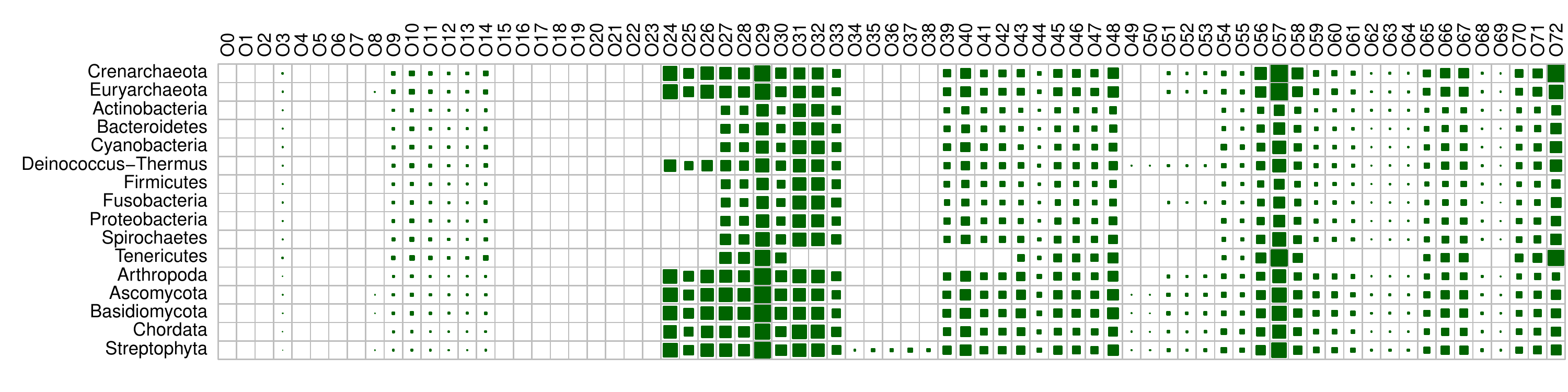}
        \caption{Differentially expressed orbits (DEO): statistically significant orbits found to have \textit{Z-score} 
$>2.58$ 
for at least $95 \%$ networks of the same phylum.}
        \label{fig:DEO}
\end{figure}

As we have discussed in the methods section orbit counts have dependencies on 
each other and we are averaging the orbit counts of a networks this may over 
represent some trivial patterns \cite{yaverouglu2014revealing}. Also, since 
OUP is a $73$ dimensional vector and all the orbit signatures may not be of 
equal evolutionary importance, some of them may be artifacts of network 
topology and their over expression may be due to shear random chance. Therefore, 
to identify the statistically significant wiring patterns we compared the OUPs 
of real PINs with their corresponding random ensembles (both ER and DDSF) by 
computing \textit{Z-scores}. Significantly overexpressed (see Methods) orbits 
having \textit{Z-score} $2.58$ or more, for at least $97\%$ of the networks 
belonging to a phyla, were considered differentially expressed and that 
particular phyla was enriched for these orbits. Differential orbit expression 
has revealed that orbits $0-2$, instances of graphlets $0$ and $1$; orbits 
$4-7$, instances of graphlets $3$ and $4$; orbits $15-23$, instances of 
graphlets $9-11$ have orbit counts comparable to their random counterparts in 
all the OUPs (Fig \ref{fig:DEO}). This suggests that these $16$ out of $73$ 
orbit topologies $(\approx 22 \%)$ are not significantly overexpressing and are 
less or not informative from evolutionary prospective. Remaining $57$ 
differentally overexpressed orbit profiles constitute our evolutionarily 
informative set of features. As observed from differentially 
expressed OUPs, within a domain of life the orbit usage is similar for most of 
phyla, however, it varies for phyla belonging to different domains of life. 
Simplest and most contrasting OUPs are of Bacterial phyla, most of them lack 
orbits \myrange[(G_{12})]{24}{26} and \myrange[(G_{20}, G_{21})]{49}{53}, 
except for Deinococcus-Thermus, that highly resemble to OUP of Streptophyta. 
Archaeal OUPs differentially express additional obits \myrange{24}{26,} 
\myrange{51}{53} and are somewhat similar to Eukaryotic OUPs which express all 
these orbits \myrange[(G_{12})]{24}{26} and \myrange[(G_{20}, G_{21})]{49}{53}. 
Orbits \myrange{34}{38}, which are instances of graphlets $15$ and $16$, are 
also not differentially expressed in any of the categories except phylum 
Streptophyta of domain Eukaryota. Mainly four orbit clusters are obtained for 
each phyla with presence or absence of some orbits in different phyla belonging 
to three domains of life resulting in a peculiar orbit expression pattern.

\subsection*{Evolutionary relevance of OUPs}

There exist several example in biology where certain vital functions are 
conserved across different species that experience greater conservative forces 
\cite{davis2015topology,kuchaiev2010topological,sharan2005conserved}. The 
proteins involved in these functions may have changes at their sequence or 
structural level in different species, however, their overall connectivity  
(wiring) pattern is same across species. For example, the core circadian clock 
proteins have variations in their sequences and structures of different 
species, however, the overall clock is preserved across different domains of 
life \cite{nohales2016molecular}, cellular information processing machineries 
\cite{kevin1999fundamentally} and metabolic processes 
\cite{peregrin2009conservation} etc. As explained in previous section the 
wiring patterns of different phyla are not obtained by random chance, so to 
further explore the biological basis of variations in the wiring patterns of 
different phyla and to know if these patterns are shaped by evolutionary 
forces, we selected five functional annotations viz. gene ontology (GO) terms, 
pathways, protein family, sub-cellular localization and protein domains to test 
the evolutionary relevance of OUPs. Pathway and protein family 
annotation data revealed that Archaeal proteins involved in differentially 
expressed orbits are participating mainly in  fundamental cellular processes 
like replication, transcription, translation, carbohydrate, lipids, amino acid, 
purine and pyrimidine metabolism. Pathways data revealed a significant 
proportion of proteins, covering on average $83 \%$ of total proteins per 
orbit, are enriching only nine fundamental Archaeal pathways, namely, 
Amino-acid biosynthesis, Cofactor biosynthesis, Purine metabolism, Pyrimidine 
metabolism, Carbohydrate degradation, Metabolic intermediate 
biosynthesis, Porphyrin-containing compound metabolism, Protein modification, 
Carbohydrate biosynthesis. Similarly most of the pathways enriched in Archaea 
are also among the most enriched pathways in Bacterial (OPAC $84 \%$) and 
Eukaryotic (OPAC $86 \%$) domains of life. Furthermore, protein family 
data also revealed most of the Archaeal, Bacterial and Eukaryotic orbit proteins 
(OPAC $82 \%, 84 \%$ and $83 \%$, respectively) are enriching families 
functioning in fundamental processes, like, transport, 
replication, transcription, translation, carbohydrate, lipids, amino acid, 
nucleic acid metabolism etc. Subcellular information revealed that $98, 85 \%$ 
of Archaeal and Bacterial orbit proteins respectively localize in only 
cytoplasm and membrane. While $69 \%$ of Eukaryotic orbit proteins localize in 
nucleus, membrane and endoplasmic reticulam membrane. This analysis also 
suggests that these orbit proteins are mostly involved in fundamental processes 
(S2). 

As observed from the pathways, protein family and sucellular location data, most 
of the orbit proteins are involved in fundamental cellular processes 
responsible for organism's growth and development. These processes are essential 
for organism's survival, and their over-all topologies (wiring patterns) remain 
conserved at large across the three domains of life 
\cite{peregrin2003phylogenetic,brasen2014carbohydrate} with subtle variations at 
molecular level \cite{kevin1999fundamentally}. In an earlier study orbit groups 
\(\{3,13, 29, 48, 55, 61\}$, $\{14, 58, 67, 71\}$, $\{72\}$, $\{4, 15, 27\}$, 
$\{10, 41, 43, 60, 64, 68\}$, $\{11, 30, 33, 42, 44\}$ and $\{12, 46, 52, 59, 
65\}\)  have been linked with processes, like, `Proteasome Assembly', 
`Transcription Initiation' and `Transcription Elongation' 
\cite{davis2015topology}. The differentially expressed OUPs obtained in this 
work, encompass most of these orbits. These consistently appearing wiring 
patterns indicate that OUPs represent topological features which are conserved 
across three domains of life and they also establish a relationship between 
network topology and function by capturing functionally similar (conserved) 
regions of different networks

\begin{figure}
     \centering
     
     \begin{subfigure}[b]{0.47\textwidth}
         \includegraphics[width=\textwidth]{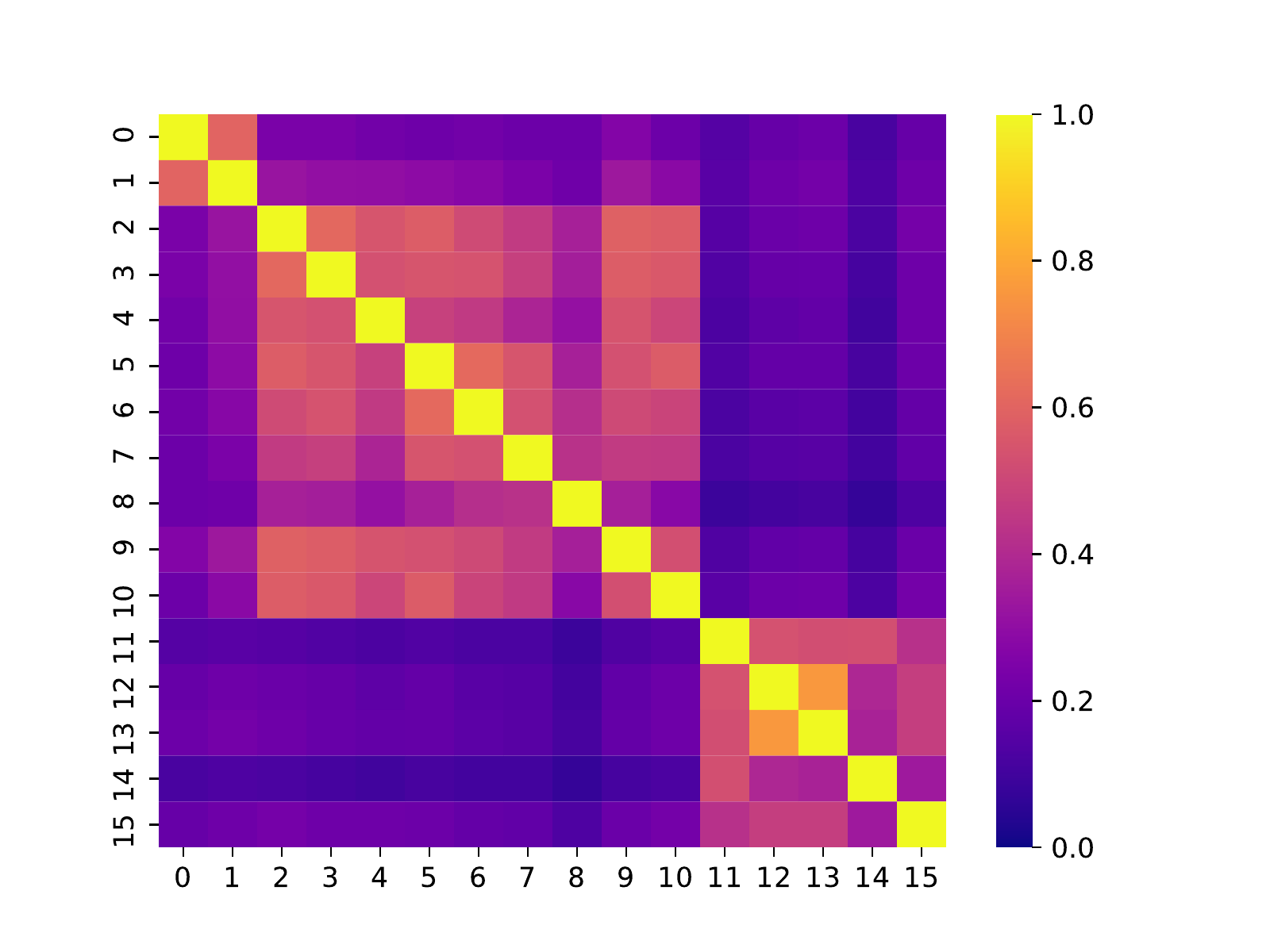}
         \caption{}
         \label{fig:PF}
     \end{subfigure}
     ~
     \begin{subfigure}[b]{0.47\textwidth}
         \includegraphics[width=\textwidth]{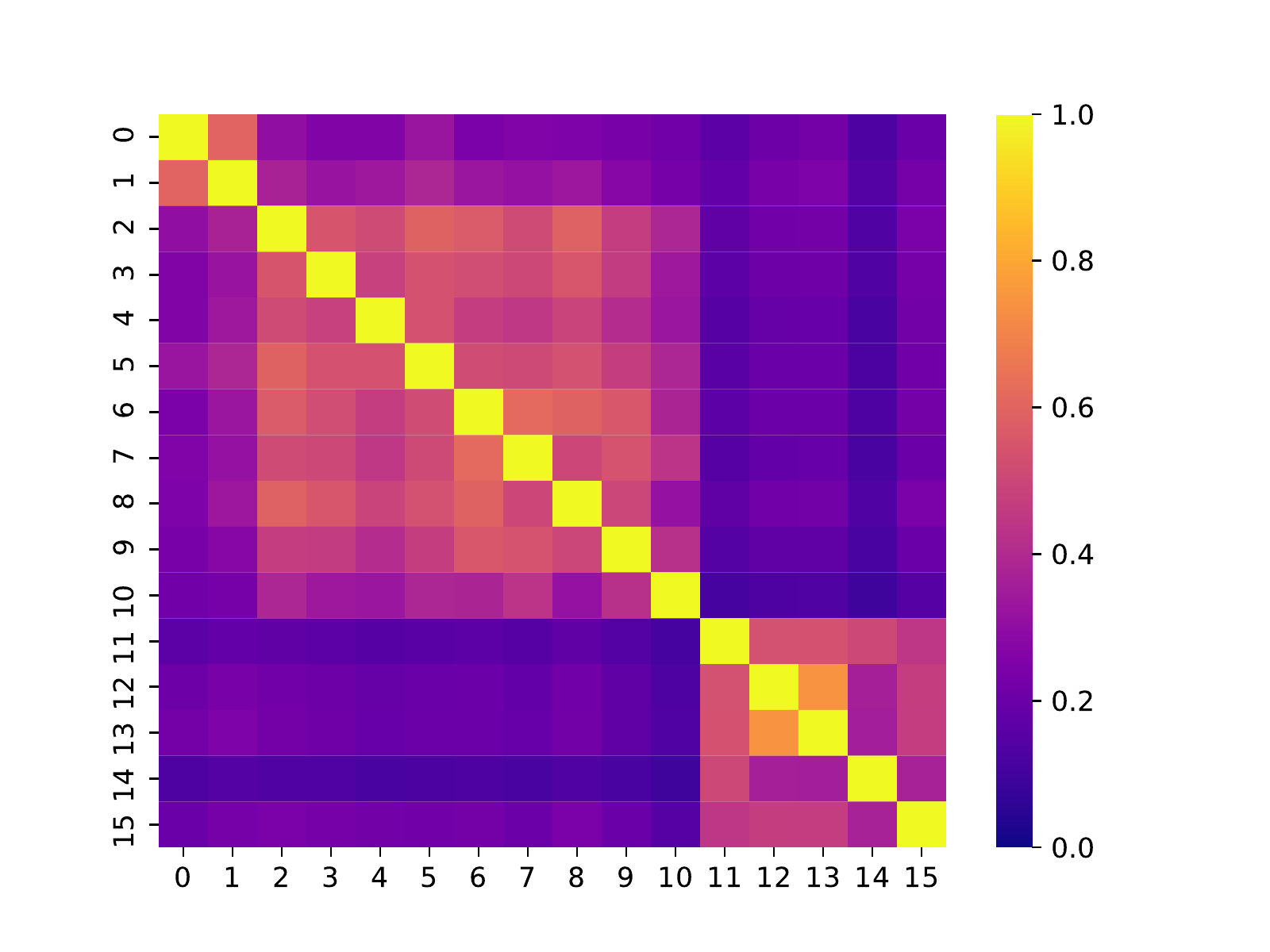}
         \caption{}
         \label{fig:DMN}
     \end{subfigure}
     
     \begin{subfigure}[b]{0.47\textwidth}
         \includegraphics[width=\textwidth]{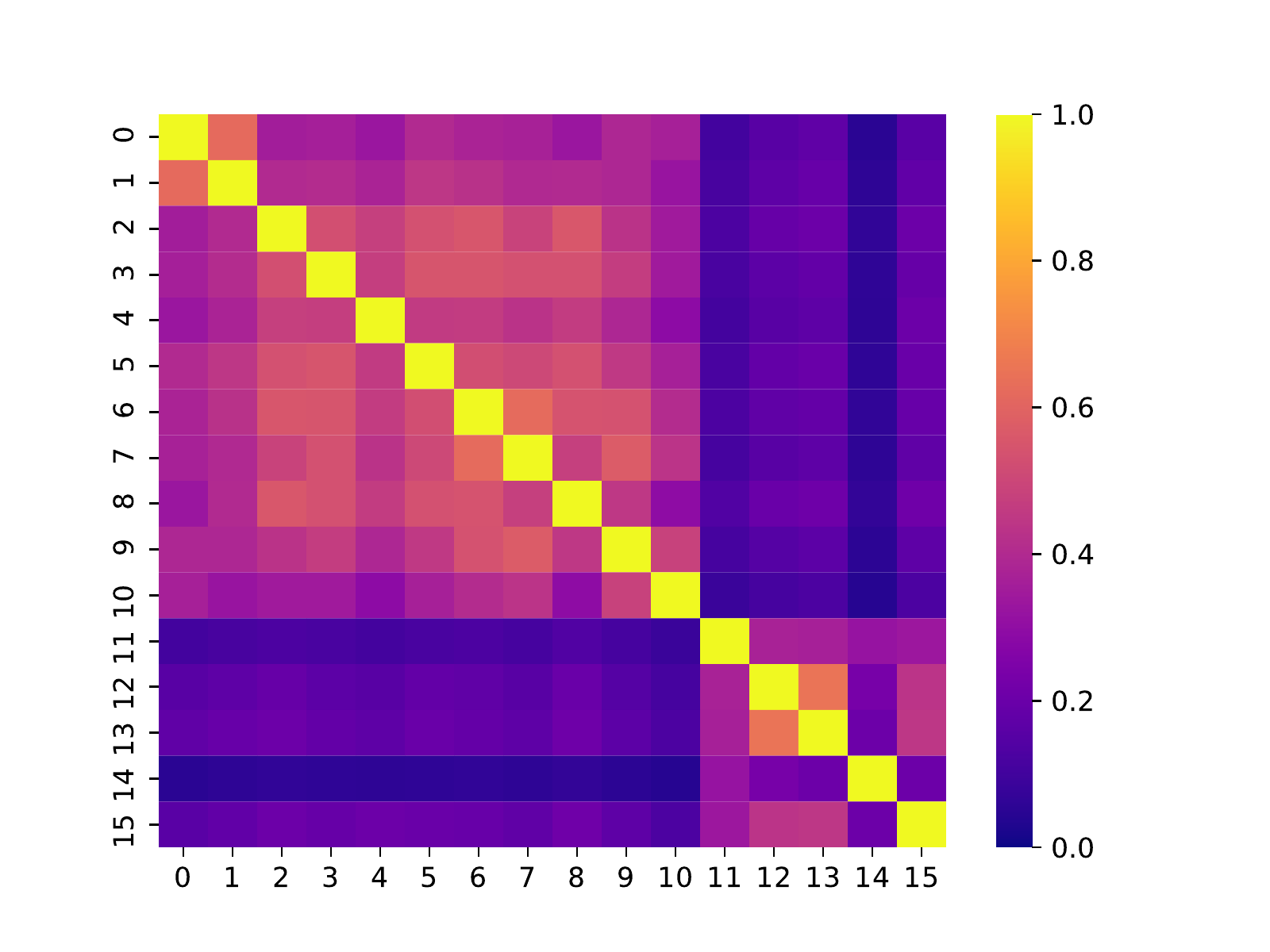}
         \caption{}
         \label{fig:GO}
    \end{subfigure}
~
     \begin{subfigure}[b]{0.47\textwidth}
         \includegraphics[width=\textwidth]{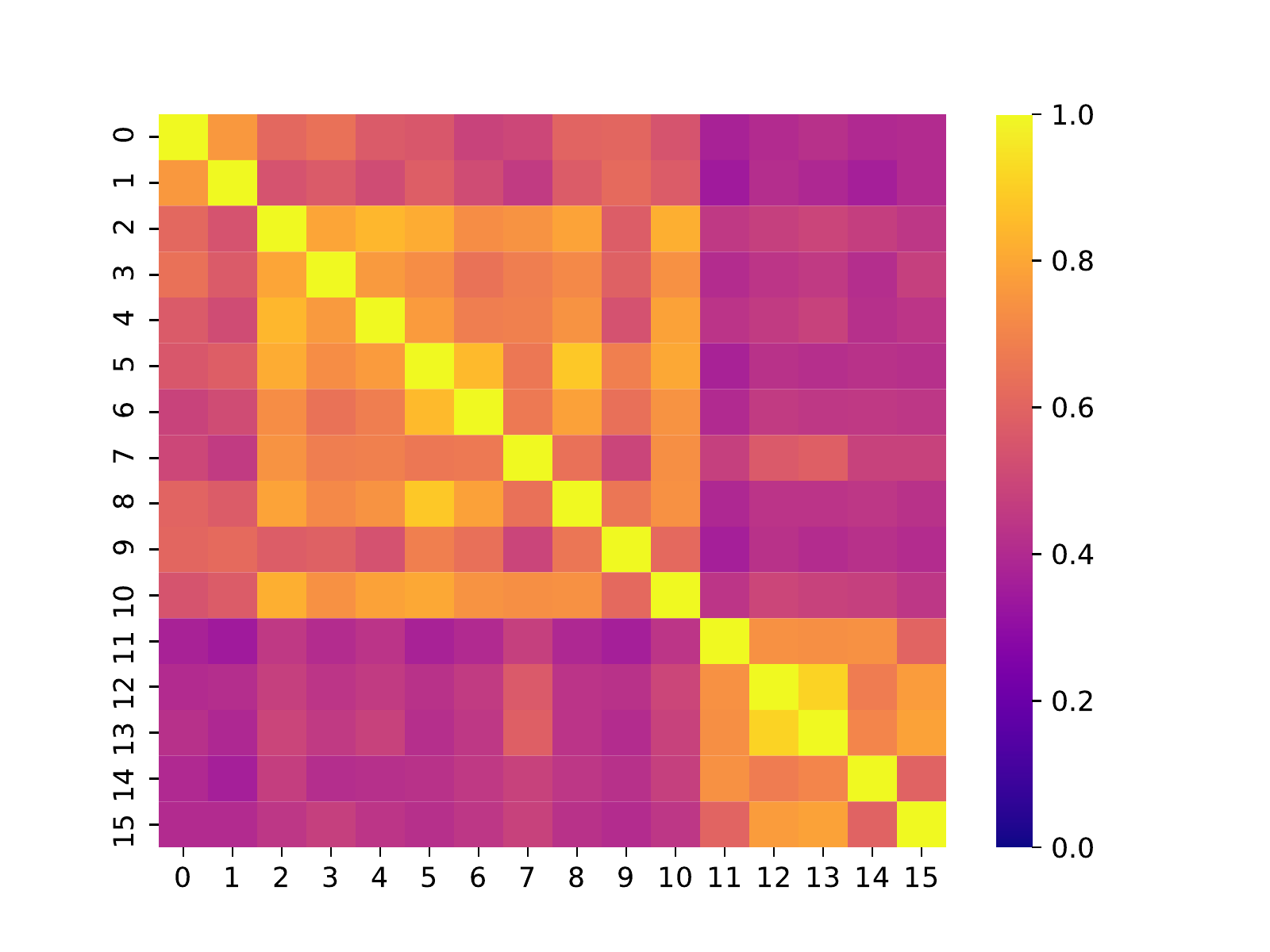}
         \caption{}
         \label{fig:PW}
     \end{subfigure}    

     \caption{Jaccard similarity (JC) matrices for function annotations, where 
each term represent an average JC value computed on pairwise JC obtained from 
$57$ orbit pairs of any two phyla: (a) JC matrix for protein family 
annotations. (b) JC matrix for protein domain annotations. (c) JC matrix for 
gene ontology annotations. (d) JC matrix for pathway annotations.}
     \label{fig:Annot}
\end{figure}

To quantify the similarity between orbit protein vectors using annotation data 
we leveraged Jaccard coefficient (JC) and constructed a pairwise similarity 
matrix among all the $16$ Phyla. Each element of this similarity matrix is an 
average JC value obtained from $57$ JC values between the two Phyla. These 
average JC matrices were computed for four functional annotations namely, GO, 
protein domains, pathways and protein family annotations. The average JC 
matrix computed from protein family annotations, covering $85 \%$ of orbit 
proteins, revealed a high average JC value $(\approx 0.6$) between the two 
Archaeal phyla Crenarchaeota and Euryarchaeota (intra-phyla) (Fig \ref{fig:PF}). 
While JC values between Crenarchaeota and other Bacterial and Eukaryotic phyla 
(inter-phyla) are comparatively very low $0.22$ and $0.16$, respectively, JC 
values among Bacterial phyla, \textit{e.g.} Actinobacteria and other bacterial 
phyla, are quite high ($0.53$) compared to Actinobacteria and other Archaeal 
($0.28$),  Eukaryotic ($0.18$) phyla. Similarily, JC values among Eukaryotic 
phyla, \textit{e.g.} Spirochaetes and other Eukaryotic phyla, are high ($0.5$) 
compared to Spirochaetes and other Archaeal ($0.15$),  Bacterial ($0.13$) phyla. 
Similarly, high JC values for intra-phyla networks and low JC values for 
inter-phyla networks are observed for protein domain annotations 
(Fig \ref{fig:DMN}). Although this distinction fades away as the scope of 
annotation is broadened. JC values computed using GO annotations are higher 
($0.61, 0.49$ and $0.35$) among intra-phyla networks (Archaea, Bacteria 
and Eukaryota), respectively, along with a comparatively higher JC values among 
Bacterial phyla and other Archaeal phyla \textit{e.g.} JC values among 
Actinobacteria and other Archaeall phyla are comparatively higher ($0.38$) than 
other Eukaryotic ($0.16$) phyla (Fig \ref{fig:GO}). This hike in JC values among 
inter-phyla networks is further increased for pathway annotations as most of 
the orbit proteins are functioning in fundamental pathways (essential) 
responsible cellular growth and development and are thus conserved across the 
three domains of life (Fig \ref{fig:PW}). These results suggest that although, 
at higher level (pathway) there are similarities among networks within a phyla, 
however, at molecular level (protein domain and family) there are subtle but 
significant enough variations in the wiring patterns that endow OUPs their 
predictive power.   

\subsection*{OUP based phylogenetic reconstruction}

\begin{figure}
     \centering
     
     \begin{subfigure}[b]{0.56\textwidth}
         \includegraphics[width=\textwidth]{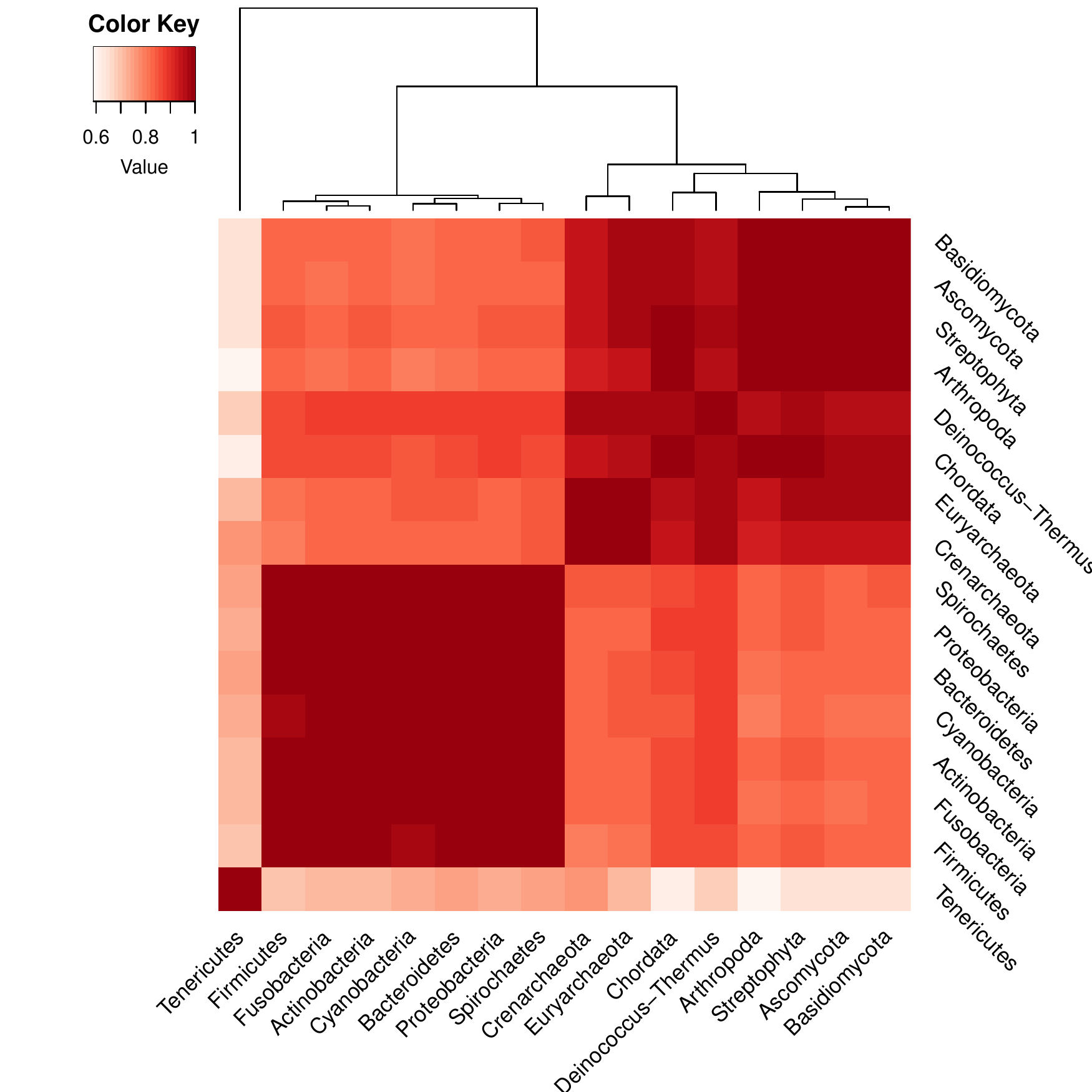}
         \caption{}
         \label{fig:DCor}
     \end{subfigure}
     
     \begin{subfigure}[b]{0.56\textwidth}
         \includegraphics[width=\textwidth]{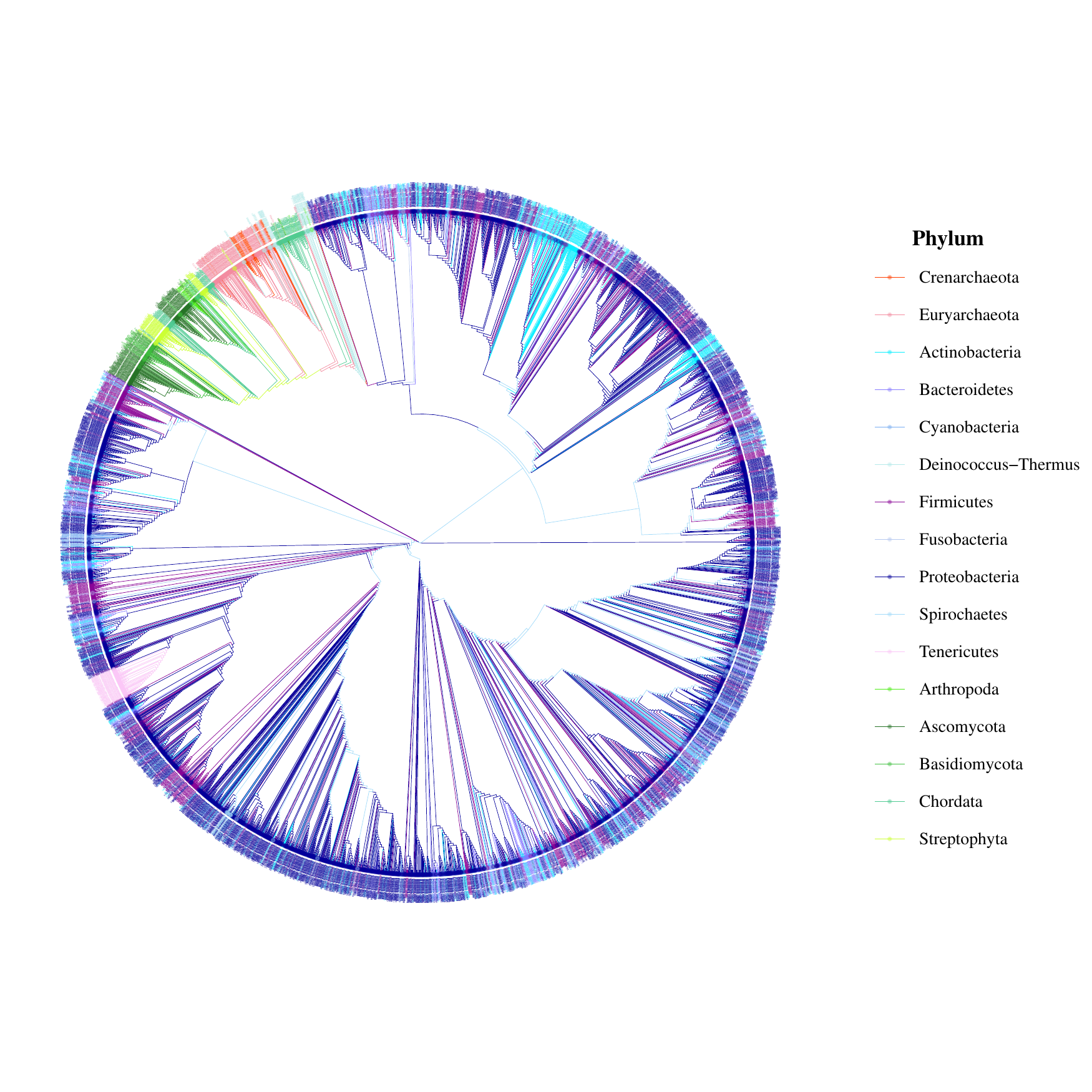}
         \caption{}
         \label{fig:HClust}
     \end{subfigure}

     \caption{Clustering of OUPs: (a) Average OUPs corresponding to each phyla 
clustered using distance correlation values. (b) All the $4,738$ OUPs clustered 
using neighbor-joining method, distance matrix computed using Bray-Curtis 
dissimilarity.}
     \label{fig:Clust}
\end{figure}

After removing non-informative orbit counts, we obtained mean OUP, by averaging 
all the OUPs of that phylum, for every phylum. Then we clustered the average 
orbit counts of all the $16$ phyla by computing pairwise distance 
correlation between them and then applying hirearchical clustering on the 
similarity matrix. We obtain three main clusters containing $1, 7$ and $8$, 
networks respectively (Fig \ref{fig:DCor}). The OUP of Bacterial phylum 
Tenericutes is found to be the most distinct from all the others and have very 
low correlation with other phyla hence placed into its individual cluster. In 
the second cluster, all but one phyla (Deinococcus-Thermus) of Bacterial domain 
are correctly grouped together in two sub clusters with Fusobacteria, 
Firmicutes and Actinobacteria being present in one sub-cluster while 
Cyanobacteria,  Bacteriodates, Proteobacteria and Sprichaetes in the other 
subcluster. In the third cluster, eight phyla are grouped together in two main 
subclusters. In one subcluster both the Archaeal phyla, Crenarchaeota 
and Euryarchaeota are grouped together, while in the  other sub-cluster 
Deinococcus-Thermus along with five Eukaryotic phyla are grouped together. Our 
method has correctly reconstructed the overall topology of the three of life 
(ToL) where the simplest Bacteria diverged earlier than the Archaeal and 
Eukryotic domains of life. However, a close observation reveals that within the 
domain, some phyla like Tenericutes, Actinobacteria, Deinococcus-Thermus are 
not at their correct positions, although some, like, Streptophyta, Ascomycota, 
Basidiomycota (all fungi) and Fusobacteria, Firmicutes, Cyanobacteria are 
correctly placed \cite{schulz2017towards}.

To know the topology of the complete tree, Bray-Curtis method of dissimilarity 
was used to compute distance matrix and the tree was constructed by 
neighbor-joining method \cite{saitou1987neighbor}. Again, similar results were 
obtained, the Archaeal phyla (band of red shades) are clustered separately and 
are close to Eukaryotic phyla (bands of green shades) which are also clustered 
separately (Fig \ref{fig:HClust}). However, the networks of bacterial phyla are 
scattered among themselves and does not form continuous blocks as for Archaea 
and Eukaryota. Both these clustering results indicate that there exists 
distinctive topological patterns conceled within OUPs and have the potential to 
reconstruct phylogenies \cite{ali2014alignment}. However, because the patterns 
are subtle they require highly sensitive methods to compute the distance. Also 
it should be noted that factors, like, noise, incompleteness of PINs etc. may 
hamper the quality of clustering \cite{yaverouglu2015proper}.

\subsection*{OUP classification}

\begin{figure}
     \centering
     
     \begin{subfigure}[b]{0.52\textwidth}
         \includegraphics[width=\textwidth]{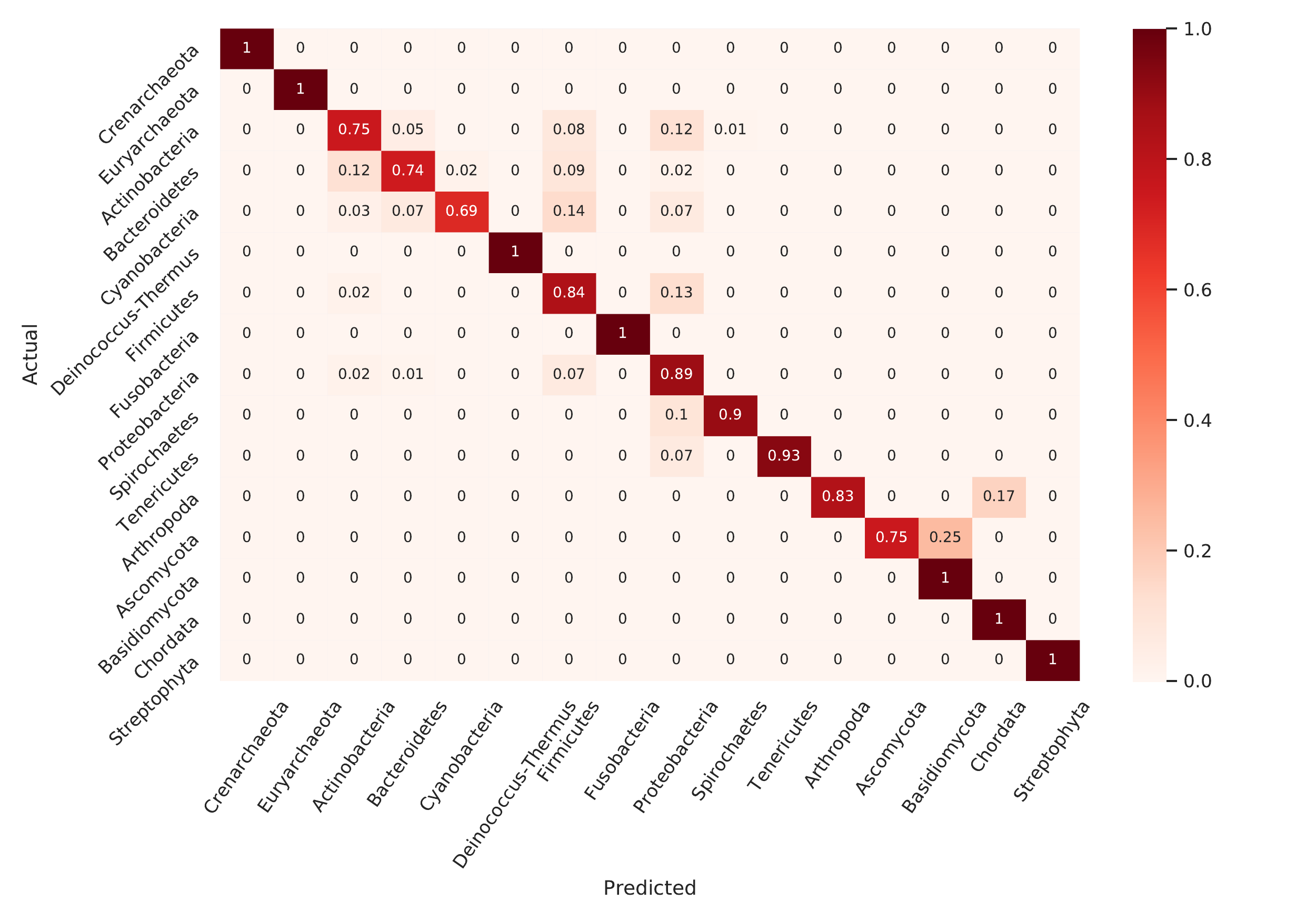}
         \caption{}
         \label{fig:ConfMt}
     \end{subfigure}
     ~
     \begin{subfigure}[b]{0.46\textwidth}
         \includegraphics[width=\textwidth]{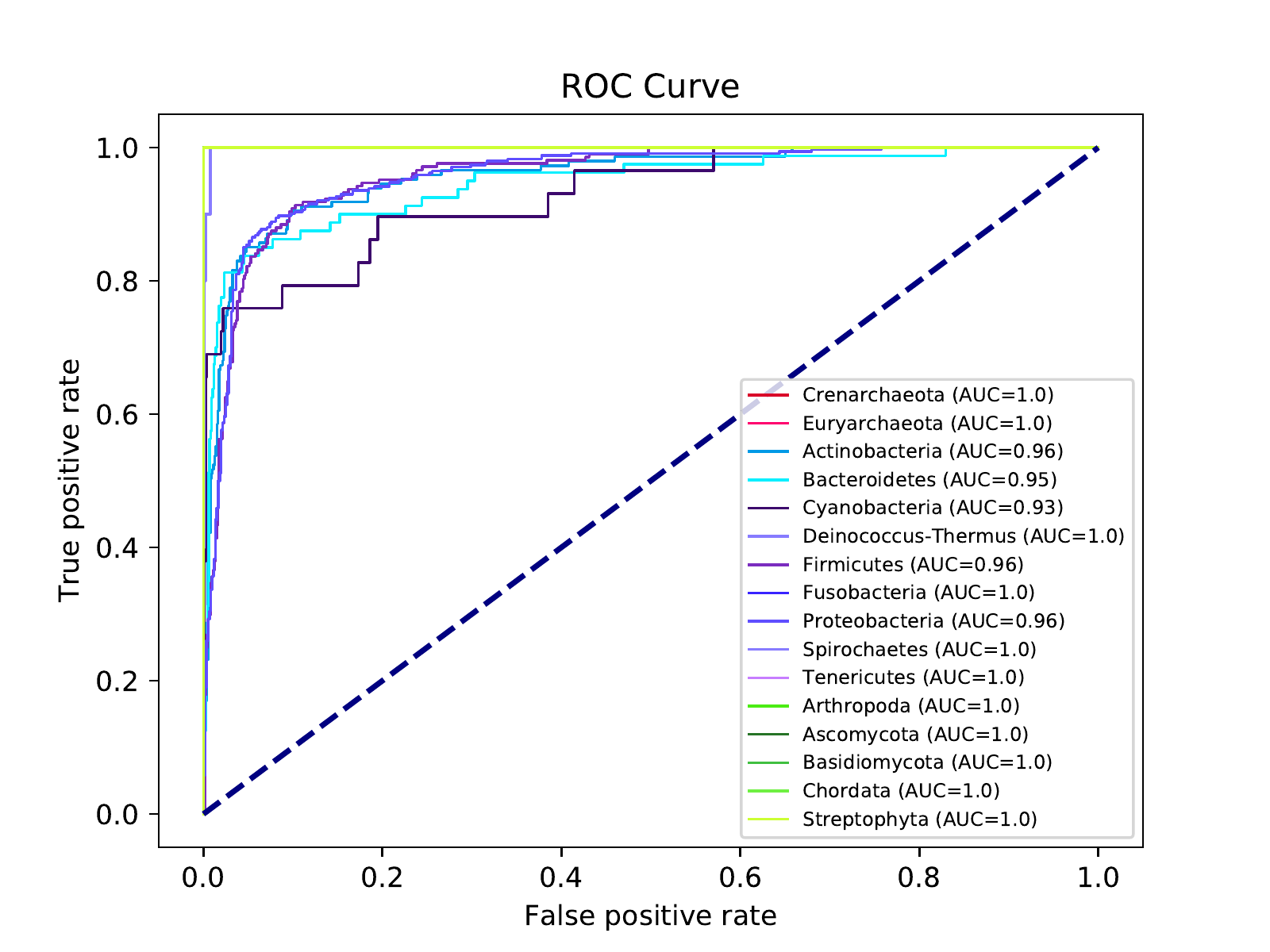}
         \caption{}
         \label{fig:ROC}
     \end{subfigure}

     \caption{Classification results of deep neural network (DeepAutoPIN) on 
$57$ dimensional OUPs: (a) Confusion matrix representing percentage of true and 
false instances predictions for every phyla. (b) ROC curves and Area under each 
ROC curve computed using one-vs-rest strategy for every phyla. }
     \label{fig:Clf}
\end{figure}

For each phylum, we learned a binary classifier using a deep neural network 
(DeepAutoPIN), where the phylum of interest represents the positive class while 
all the other phyla correspond to the negative class, to predict the phylum of 
a PIN with an unknown phylum label. To evaluate the performance of the deep 
neural network, for each binary classification we computed Area Under the ROC 
Curve (AUC) for every classifier and also developed a confusion matrix. 
Classification results are presented as confusion matrix (Fig \ref{fig:ConfMt}), 
that revealed the presence of distinct graphlet wiring patterns specific to 
phyla as well as the three domains of life. Our classifier successfully 
differentiated among networks belonging Archaeal, Bacterial and Eukaryotic 
domains without any misclassification. Our model successfully classified all 
the instances of Archaeal networks, both Crenarchaeota and Euryarchaeota, into 
their respective phyla with an AUC value of \(1.0\), indicating that they 
possess signatures which are distinct from the networks belonging to Bacterial 
and Eukaryotic phyla as well as from each other. This is consistent with 
results obtained from distance correlation where we found that OUPs of both the 
Archaeal phyla are highly correlated with each other. Similarly, our 
model successfully classified all the networks of Basidiomycota, Chordata and 
Streptophyta into their respective phyla. Although on networks belonging to 
Arthropoda and Ascomycota, our model achieved $83 \%$ and $75\%$ accuracy 
respectively. In case of Ascomycota $25 \%$ of networks are mislabeled as 
Basidiomycota (both of these are fungi), while $17 \%$ of Arthropoda networks 
are mislabeled as Chordata (both of the phyla belong to animal kingdom).  For 
all the five Eukaryotic domains the AUC score was observed to be \(1.0\). 
Similar to both Archaeal and Eukaryotic networks, OUPs of Bacterial networks 
possess signatures distinct enough to classify them as being Bacterial. Our 
model correctly predicts all the networks of Deinococcus-Thermus and 
Fusobacteria, however, it achieved $75 \%, 74 \%, 69 \%$ classification accuracy 
on OUPs obtained from Actinobacteria, Bacteriodates and Cyanobacteria 
respectively. While OUPs of Firmicutes, Proteobacteria, Spirochaetess and 
Tenericutes were classified with corresponding prediction proabilities $84 \%, 
89 \%, 90 \%, 93 \%$, respectively.  Phylum Actinobacteria, Bacteriodates and 
Cyanobacteria are among the top three bacterial phyla for which mislabelling 
rate is highest followed by Firmicutes and Proteobacteria, however, most of the 
mislabeled networks are classified into Firmicutes or Proteobacteria. This 
highlights that these two bacterial phyla have most general topological 
features which are shared between most of the Bacterial phyla considered in 
this study. AUC score for all of the $16$ classes are very high (closer to 1) 
indicating a high predictive performance of our model (Fig \ref{fig:ROC}). Our 
model obtained an overall prediction accuracy of $85 \%$ on $16$ class test set.

\section*{Summary and Conclusions}

We propose an approach to compare PINs belonging to $16$ different phyla by 
summarizing the subtle wiring patterns called  orbit usage profiles (OUPs). 
Diffferentially expressed orbits profiles characterized by $57$ evolutionarily 
informative orbits were obtained and found that there exists phyla specific 
topological features conserved across the three domains of life. These profiles 
are similar for within phyla PINs (intra-phyla) and are distinct from the PINs 
belonging to other phyla (inter-phyla). Our method is able to capture conserved 
biological processes underlying PINs, using data of five functional annotations 
a relationship between the topological features of orbit proteins and 
biological functions is establish and found that the topological features 
represent functions vital for sustenance of life that are shaped by 
evolutionary forces acting on them. Furthermore, we have shown that our approach 
is able to group networks into their respective phyla. Notably a deep neural 
network (DeepAutoPIN) trained on OUPs achieved an overall accuracy 
of $85 \%$ to predict the label of a PIN that has not been assigned to any 
phylum. Most of the phyla contain distinct topological features from other 
phyla, however, there exists some indistinguishable features among Eukaryotic 
phyla as well as Prokaryotic phyla. To the best of our knowledge most of the 
studies on PINs have been limited to organisms with high PPI coverage, this is 
the first study that systematically investigate $4,738$ PINs and highlights 
evolutionarily conserved topological features that shaped their overall 
architecture across the tree of life. Since the available interactomes are 
sparse and contain a substantial number of false negative and positive 
interactions, a considerable limitation, despite consistent results our 
findings may become more generalized with increasing interactomic data.

\section*{Acknowledgements}
$\text{VS}^{\dagger}$ thanks Council of Scientific
and Industrial Research (CSIR), India for providing Senior Research Fellowship
(SRF). \textbf{Funding:} Authors received no specific funding for
this research work. \textbf{Authors Contributions:} $\text{VS}^*$
conceptualized and designed the research framework. $\text{VS}^{\dagger}$
performed the computational experiments. $\text{VS}^{\dagger}$ and
$\text{VS}^*$ analyzed the data and interpreted results. $\text{VS}^{\dagger}$
and $\text{VS}^*$ wrote and finalized the manuscript. \textbf{Competing
Interests:} The authors declare that they have no conflict of interests.
\textbf{Data and materials availability:} Supplementary data is available
at \url{https://drive.google.com/file/d/1TkiW0bKe0sSEiFm3bREEWP_M0R2iitmS/view?usp=sharing} while the code for DDSF is available
at \url{https://github.com/vikramsinghlab/DDSF}, and for DeepAutoPin is
available at \url{https://github.com/vikramsinghlab/DeepAutoPIN}.
\textbf{Ethics approval: } Not applicable. \textbf{Concent to participate: }
Not applicable. \textbf{Concent to publish: } Not applicable.

\pagebreak

\bibliography{P1}
\bibliographystyle{unsrt}

\pagebreak

\begin{algorithm}
\begin{codebox}
\Procname{\(\proc{DDSF}(n, m, \langle k \rangle)\)}
\li    \Comment Generates a density dependent scale free random graph with 
\li    \Comment n nodes and \(\approx\) m edges having average degree $\langle k \rangle$  
\li    \(A \gets (0, 1)\)
\li    \(\id{N_{current} \gets 2} \)
\li    \(\id{New-links} \gets 0\)
\li    \Comment Seed generation
\li    \While \(\id{N_{current}} \leq \langle k \rangle\)
\li    \Do
            \For \(i \gets 0 \To \id{N_{current}} - 1\)
\li        \Do
                 \(r \gets \id{int}(\id{rand(\attrib{A}{length}))}\)
\li             \(\attrib{A}{add\_edge} \gets (\id{N_{current}}, A[r])\)
            \End
\li        \(\id{N_{current}} \gets \id{N_{current}} + 1\)
        \End

\li    \Comment Extending the seed
\li    \While \(\id{N_{current}} < n\)
\li    \Do
            \(\langle k_{rem} \rangle \gets \frac{2 \times (m - \frac{\attrib{A}{length}}{2})}{n - \id{N_{current}}}\) 
\li        \(k \gets \id{int}(\langle k_{rem} \rangle)\)
\li        \If \(\langle k_{rem} \rangle < 1\)
\li            \Then
                    \(\id{New-links} \gets 1\)
\li            \Else
\li                \(\id{r_1} \gets \id{rand}(1)\)
\li                \If \(\id{r_1} < \langle k_{rem} \rangle - k\)
\li                    \Then
                            \(\id{New-links} \gets k\)
\li                    \Else
\li                        \(\id{New-links} \gets k+ 1\)
                        \End
                \End
\li        \(\id{N_{current}} \gets \id{N_{current}} + 1\) 
            \End
\end{codebox}
\caption{Density dependent scale free graph}
\label{Algo:A1}
\end{algorithm}
\end{document}